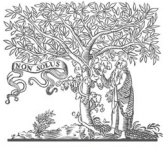
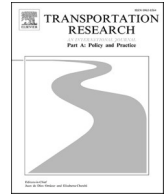

# Traveller behaviour in public transport in the early stages of the COVID-19 pandemic in the Netherlands

Sanmay Shelat [a,*], Oded Cats [a], Sander van Cranenburgh [b]

[a] *Department of Transport and Planning, Delft University of Technology, Netherlands*
[b] *Transport and Logistics Group, Delft University of Technology, Netherlands*

ABSTRACT

With a few exceptions, public transport ridership around the world has been hit hard by the COVID-19 pandemic. Travellers are now likely to adapt their behaviour with a focus on factors that contribute to the risk of COVID-19 transmission. Given the unprecedented spatial and temporal scale of this crisis, these changes in behaviour may even be sustained after the pandemic. To evaluate travellers' behaviour in public transport networks during these times and assess how they will respond to future changes in the pandemic, we conduct a stated choice experiment with train travellers in the Netherlands at the end of the first infection wave. We specifically assess behaviour related to three criteria affecting the risk of COVID-19 transmission: (i) crowding, (ii) exposure duration, and (iii) prevalent infection rate. Observed choices are analysed using a latent class choice model which reveals two, nearly equally sized latent traveller segments: 'COVID Conscious' and 'Infection Indifferent'. The former has a significantly higher valuation of crowding, accepting, on average 8.75 min extra waiting time (average total travel time in the choice scenarios was about 40 min) to reduce one person on-board. Moreover, this class indicates a strong desire to sit without anybody in the neighbouring seat and is quite sensitive to changes in the prevalent infection rate. By contrast, the Infection Indifferent class has a value of crowding (1.04 waiting time minutes/person) that is only slightly higher than pre-pandemic estimates and is relatively unaffected by infection rates. We find that older and female travellers are more likely to be COVD Conscious while those reporting to use the trains more frequently during the pandemic tend to be Infection Indifferent. Further analysis also reveals differences between the two segments in attitudes towards the pandemic and self-reported rule-following behaviour. We believe that the behavioural insights from this study will not only contribute to better demand forecasting for service planning but will also inform public transport policy decisions aimed at curbing the shift to private modes.

## 1. Introduction

The COVID-19 pandemic has led to unprecedented restrictions on public life globally. Some of the first restrictions in many places were on public transport which, by its very nature of moving people in dense, enclosed spaces, could be a major transmission risk for this highly contagious virus[1]. While some public authorities completely stopped service (e.g., India (Union Home Secretary, 2020)), others restricted or discouraged use other than by essential workers or for urgent needs (e.g., Netherlands, United Kingdom (Department for Transport; Openbaar Vervoer Nederland, 2020a)). Then, at the end of the first infection wave, many authorities, pressed with a need to restart economies and provide essential transportation, eased restrictions and cautiously resumed public transport. Demand levels, however, did not return to pre-pandemic levels (Citymapper, 2021; Google LLC, 2021), at least partly, due to heightened (awareness of the) risk of infection (Beck and Hensher, 2020).

* Corresponding author.
 *E-mail addresses:* S.Shelat@tudelft.nl (S. Shelat), O.Cats@tudelft.nl (O. Cats), S.vanCranenburgh@tudelft.nl (S. van Cranenburgh).

The effect of on-board crowding on travel behaviour has received much attention in literature and has been widely accepted to be a significant influence on various choice dimensions (Hensher et al., 2011; Li and Hensher, 2011; Tirachini et al., 2013; Wardman and Whelan, 2011). Using choice observations, mainly from stated choice experiments (e.g., (Kroes et al., 2013; Sahu et al., 2018)) but also from revealed preferences (e.g., (Hörcher et al., 2017; Yap et al., 2018)), a number of studies have estimated the value of crowding in terms of the willingness to pay to reduce it or its impact on the value of travel time. The disutility of crowding in these studies arises primarily from physical and psychological discomfort and exhaustion. However, given the wide and sustained impact of the COVID-19 pandemic, travellers are now likely to want to avoid crowds even more so than under normal circumstances as a measure towards minimizing their exposure to the virus (Tirachini and Cats, 2020).

Travellers may now focus on factors contributing to COVID-19 transmission and for service planners to be able to respond to these changes in behaviour it is essential to have an empirical underpinning of those. The question is then: given the COVID-19 pandemic, how will travellers respond to crowdedness on public transport vehicles and future changes in infection rates? Studies on the COVID-19 pandemic as well as those on previous epidemics resulting from viruses spread through similar means (such as SARS, MERS, swine flu) have shown that people perceive avoiding public transport as a preventive measure (Gerhold, 2020; Kim et al., 2017; Lau et al., 2003; Rubin et al., 2009). A number of COVID-19 related analyses also indicate a significant mode shift to private modes such as bicycles and automobiles demand (e.g., Bucsky (2020)). While these studies focus on perceptions and aggregate statistics, only a few studies have analysed public transport travellers' choice behaviour.

A Scopus search[2] and other modes of literature collection found only a handful of studies conducting choice analysis in the context of public transport and epidemics. Scorrano and Danielis (2021) conduct a mode choice analysis for before and during COVID-19 in Trieste, Italy. The impact of the pandemic on mode choice is parametrised as mode-specific penalties which they find to be negative (and even more so for COVID-19 risk averse travellers) for public transport. Cho and Park (2021) and Aghabayk et al. (2021) also conduct a before-after experiment but focus on estimating crowding multipliers using stated choice data from Seoul and Tehran, respectively. They find that crowding multipliers are 1.04–1.44 times higher during the pandemic, confirming expectations that travellers would be more wary of crowds. Awad-Núñez et al. (2021) and Aaditya and Rahul (2021) focus on the impact of COVID-19 safety measures (such as reducing on-board crowding and improving cleaning) on the willingness to use and pay for public transport. They find that a higher safety perception increased willingness to use public transport. Finally, Hensher et al. (2022) present choice models of commuting, work-from-home, or not working using revealed choice data from Sydney and South East Queensland.

We contribute to the growing literature on COVID-19 and public transport by analysing how travellers have adapted their behaviour under these exigent circumstances. A stated choice experiment is conducted to analyse traveller behaviour specifically related to three criteria affecting the risk of COVID-19 transmission (Hu et al., 2020; Prather et al., 2020): (i) distance to other people, (ii) duration of exposure, and (iii) prevalent infection rate. In the context of public transport travel, the first two correspond to on-board crowding and in-vehicle time, respectively. With the choice experiment, we measure travellers' crowding valuation in the backdrop of the ongoing pandemic and how these valuations are affected by factors that might affect the perception of related risk. These model estimates will not only be useful for demand forecasting but could also provide insights that may be valuable for policy designs aimed at managing demand (Gkiotsalitis and Cats, 2020) not only for the ongoing crisis but also the next pandemic. In this study, we report findings from the stated choice experiment conducted with train travellers in the Netherlands towards the end of the first infection wave, just as the first restrictions were being lifted.

In the next section, we describe the survey design, data collection, and choice analysis methodology. This is followed by the results and discussions in Section 3. Finally, a summary of the results, potential policy implications, limitations of the study, and future avenues of research are outlined in Section 4.

## 2. Data and analysis

To understand traveller behaviour under the new circumstances presented by the pandemic, a stated choice experiment was conducted with Dutch train travellers. The experiment was part of a larger survey that collected, among other things, travellers' socio-demographics, mobility choices, and pandemic-related qualitative measures. Discrete choice analysis is applied on observations from the experiment to measure crowding valuation while the personal characteristics are used to explain heterogeneity in behaviour either *a posteriori* or as part of the choice model.

---

[1] However, there is no conclusive evidence to this end and indeed some suggest that if recommended mitigation measures are implemented, the risk of contracting COVID-19 in public transport could be low (Gkiotsalitis and Cats, 2020; Goldbaum, 2020).

[2] Scopus search term (initial search on 25 March 2021 revealed only two relevant studies; the overview was updated to briefly include studies published during the review based on a search on 25 December 2021):(TITLE-ABS-KEY (pandemic OR epidemic OR sars OR mers OR "swine flu" OR h1n1 OR ebola OR covid) AND TITLE-ABS-KEY ("public transport*" OR transit OR bus OR tram OR train OR metro) AND TITLE-ABS-KEY ((choice OR logit OR probit) W/2 (model* OR analys*)))

## 2.1. Survey design

### 2.1.1. Stated choice experiment

The experiment consists of a series of choice situations in which respondents were asked to assume that they had arrived at a train station from which two trains were available for their destination. They were informed that they were travelling with the same purpose for which they had indicated they most frequently used the train before the pandemic-related restrictions. The train alternatives varied only in terms of on-board crowdedness (distance to other people) and waiting time. We note that this means that crowding valuation will be obtained in terms of waiting time savings instead of the usual money amounts. We did not use different travel costs directly to avoid interactions with respondent income and expectations from a higher travel class. Implying that a more costly train would be less crowded (and therefore safer) could also lead to protest answers. Contextual information about factors potentially affecting the perception of contracting the disease, namely travel time (exposure duration) in either train, and prevalent infection rate, was also given. The latter was provided in terms of the proportion of Dutch population that is infectious and capable of transmitting the virus to others. To ensure that only in-vehicle time was considered as the duration of exposure, we noted that it was possible to maintain social distance while waiting. Furthermore, respondents were reminded of the mandatory face mask regulations on-board public transport vehicles (Openbaar Vervoer Nederland, 2020b). While this was not mentioned explicitly in the survey, we note that in majority of the trains, windows cannot be opened, meaning that ventilation conditions cannot be easily changed. Respondents were asked to rank the two train alternatives and the option of not travelling by train for each choice situation. We asked for a ranking rather than a single best choice to enable us to obtain trade-off estimates in the case that the majority of respondents chose to opt-out altogether.

On-board crowding was presented graphically as the seated section in a single coach of a commuter train (known as *Sprinter* in the Netherlands). Five levels of crowdedness were used: 5, 18, 23, 28, and 36 seats occupied (out of 40), colloquially corresponding to the following labels: 'almost empty', 'able to sit alone', 'unable to sit alone but not too crowded', 'quite crowded', and 'packed' (Fig. 1). To avoid confusion, respondents were informed that the indicated crowding level was after everyone else (but the respondent) had boarded. While these trains do have some standing space near the doors (not shown in the graphics), we excluded the option to stand in order to simplify the choice situation. By only offering seating space, we may miss out on capturing the (possibly different) crowding valuation of those who would prefer to stand. However, given the limited and confined standing space, and the relatively long in-vehicle time levels used in the study (see below), it is likely that most travellers would have preferred to find a seat for their trip. As such, we expect the impact of this simplification to be small. To trigger respondents to consider where they would sit, they were asked, in a series of questions prior to the choice experiment, to indicate where they would sit in each of the five crowding levels. Three levels of waiting times were used: 3, 12, and 25 min. A wide range was deliberately used to ensure that we would observe trade-offs between on-board crowding and waiting time savings.

It is likely that respondents would find it difficult to respond to infection rate numbers without any real-world references on which to anchor their evaluation of this variable. To help respondents interpret the infection rate numbers, we sought to provide them a best estimate of the infection levels (i) at the time of the survey when restrictions had begun to be lifted (0.1%) and (ii) at the peak of the pandemic (in terms of daily reported cases and hospitalizations) in mid-April (0.43%). The proportion of infectious people in the population is innately unknowable due to the presence of asymptomatic and pre-symptomatic cases, limited testing capacity, and reluctance to get tested. Therefore, in the absence of official estimates (at the time of the survey[3]), we obtained the above numbers from back-of-the-envelope calculations using daily reported infections. In the experiment, five levels around these reference infection rates were used: 0.01% (pre-restriction levels), 0.1% (at the time of the survey), 0.5% (mid-April level), 2%, and 10% (extremely high). For the other contextual variable, in-vehicle time, three levels were used: 10, 25, and 40 min.

We used a semi-random experiment design: weakly dominated and symmetrical choice situations were removed from the full factorial of the above described attribute levels; and from these, 4 subsets of 15 choice situations were randomly picked. Respondents then faced one of these subsets at random. Walker et al. (2018) argue that semi-random designs, where dominated choice tasks are eliminated, perform as well as efficient designs, particularly because they are robust against a large range of parameter estimates and model specifications. A screenshot of the experiment is shown in Fig. 2.

### 2.1.2. Personal characteristics

Three categories of personal characteristics were collected to explain potential differences in behaviour: (i) mobility factors, (ii) socio-demographic factors, and (iii) COVID-19-related qualitative measures. In the first category, we asked travellers how often they travelled with the train before and during the pandemic-related restrictions, the crowding level they usually experienced, their most frequent purpose of travel, and which alternative modes were available for these trips. In the second category, common socio-demographic questions (age, gender, income, employment status, zip code, highest education attained, and household size) were asked. In addition, some variables more specific to the current context were also collected; in particular, ages of household members and past, current, and expected future status of working or studying from home. The final category consists of questions regarding the perceived likelihood of the respondent or someone in their household getting infected and the severity of the disease if they do. Respondents were also asked about the degree to which they think they, themselves, and others follow pandemic related advice and regulations such as frequent hand sanitization and social distancing in public places. Finally, this category also includes questions about institutional trust and frequency of information seeking in relation to the pandemic. Note that all variables in this category except the last one noted here are qualitative Likert scale measures.

---

[3] Since then, the Dutch government has published these figures (also retroactively) (Rijksoverheid, 2021). Their estimates for April 15 and May 20 (at the time of the survey) are 0.34% and 0.08%, respectively. Although these values are fairly close to ours, they estimate the peak of the infection rate to be around the end of March rather than mid-April.

| Crowding Level | Graphic |
|---|---|
| Almost empty | 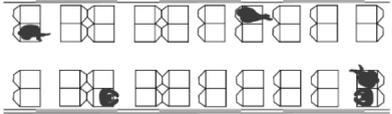 |
| Able to sit alone | 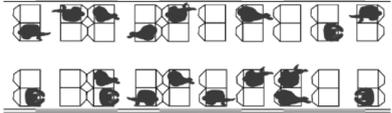 |
| Unable to sit alone but not too crowded | 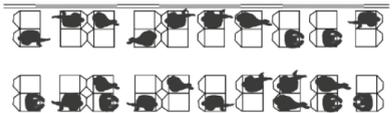 |
| Quite crowded | 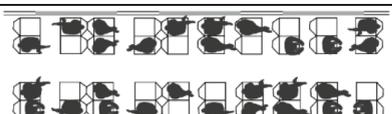 |
| Packed | 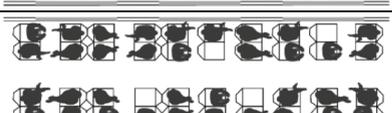 |

**Fig. 1.** Graphical presentation of crowding levels.

*2.2. Data collection*

The survey was distributed to Dutch train travellers who travelled by this mode at least once per month before March 2020 when the first pandemic-related restrictions were imposed. In March 2020, the Dutch government urged travellers to use public transport only 'if it is really needed'. By May 2020, having achieved a reduction in the daily reported cases, Dutch authorities announced that certain professions, services, and educational activities could resume by the end of that month (Rijksoverheid, 2020). Furthermore, public transport could be used once again by mid-June 2020 but with new regulations such as mandatory face masks and seat blocking to maintain distance (the latter was stopped in July 2020) (Openbaar Vervoer Nederland, 2020b). Data collection took place from 20 to 25 May, after announcements concerning these measures had been made. A total of 513 valid responses[4] were collected via an online panel. The survey was offered in Dutch and we expected a completion time of 12–15 min. In addition to pre-COVID-19 train use requirements, we sought to collect a sample representative of the overall Dutch population in terms of age, gender, and education level. Table 1 shows some sample characteristics.

*2.3. Choice analysis*

Observations are analysed under the conventional random utility maximization framework where the utility of an alternative $i$ for individual $n$, $U_{in}$, consists of a systematic ($V_{in}$) component, capturing the utility associated with factors observed by the analyst, and a random ($\varepsilon_{in}$) component. We assume that the systematic component is linear-additive and is computed by taking the sum of the alternate specific constant ($\beta_i$) and the product of taste preferences ($\beta_{ij}$) and the values of attributes, $j$ ($x_{ijn}$) (Equation (1)). The probability of choosing alternative $i$ from $I$ alternatives in a multinomial logit (MNL) model, is given by Equation (2).

$$U_{in} = V_{in} + \varepsilon_{in}$$
$$V_{in} = \beta_i + \sum_j \beta_{ij} \cdot x_{ijn} \tag{1}$$

$$P_{in} = \frac{e^{V_{in}}}{\sum_{i'=1}^{I} e^{V_{i'n}}} \tag{2}$$

---

[4] About 40 responses completed in<6 min were removed as this was considered to be too fast to have been properly answered. Response time was not a significant indicator in a simple linear-additive multinomial logit model and a model with just these responses returned many insignificant (p < 0.05) parameters indicating randomness in the responses given.

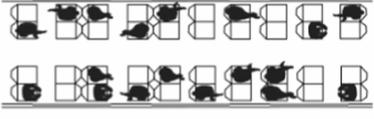

Fig. 2. Screenshot of the choice experiment (translated to English).

To assess heterogeneity in traveller behaviour we use a latent class choice model (LCCM) which is a discrete mixture of choice models to which individuals are probabilistically allocated. Although the choice models can have different attributes, structures, or even belong to a completely different framework, we use the same model in each class. The probability of an individual *n*, belonging to class *s* (amongst S classes) with probability $\pi_{ns}$, choosing alternative *i* is the product-sum of the class membership probabilities and the probability of selecting that alternative for each class (given the vector of taste parameters in that class, $\beta_s$) (Equation (3)). Panel effects are accounted for by assuming that a particular individual is allocated to each class with the same probability for all their choices. The likelihood of observing an individual's sequence of choices $i:i_1,...,i_T$ by individual *n* over *T* situations is given by Equation (4).

$$P_{in} = \sum_{s=1}^{S} \pi_{ns} \cdot P_{in}(\beta_s) \qquad (3)$$

$$L_{in} = \sum_{s=1}^{S} \pi_{ns} \prod_{t=1}^{T} P_{in_t}(\beta_s) \qquad (4)$$

An important aspect of LCCM is the ability to explain behavioural heterogeneity though class membership probabilities using values of individual characteristics, $k$ ($z_{kn}$) (Equation (5)). We use socio-demographic and mobility characteristics to explain class membership. For other, generally unobservable variables (such as, worrying about transmitting the infection to someone in the household), we conduct a posterior analysis to find the distributions of these variables in the classes of the estimated model.

$$\pi_{ns} = \frac{e^{\delta_s + \sum_k \gamma_{ks} \cdot z_{kn}}}{\sum_{s'=1}^{S} e^{\delta_{s'} + \sum_{k'} \gamma_{k's'} \cdot z_{k'n}}} \qquad (5)$$

**Table 1**
Sample characteristics.

| Total respondents | | 513 | |
|---|---|---|---|
| | | **Distribution (%)** | |
| **Attribute** | **Value** | **Actual** | **Required**[1] |
| Gender | Female | 49% | ~50% |
| | Male | 50% | ~50% |
| | Other | 0% | |
| Age | 18–24 | 15% | 11% |
| | 25–34 | 18% | 16% |
| | 35–44 | 17% | 15% |
| | 45–54 | 17% | 18% |
| | 55–64 | 19% | 17% |
| | 65–74 | 13% | 14% |
| | >74 | 2 %[2] | 10% |
| Education[3] | Elementary school (*basisonderwijs*) | 1% | |
| | Secondary school (*HAVO/VWO/VMBO*) | 27% | ~29% |
| | Vocational diploma (*MBO*) | 34% | ~37% |
| | Higher professional education (*HBO*) | 25% | |
| | University education incl. bachelor, master, PhD (*WO*) | 13% | ~33% |
| Trip purpose | Work | | 38% |
| | Education | | 11% |
| | Visiting friends or family | | 26% |
| | Recreation (e.g., sightseeing) | | 17% |
| | Errands (e.g., supermarket, bank, etc.) | | 1% |
| | Business visit | | 6% |

[1] Source: Centraal Bureau voor de Statistiek (2020).
[2] The under-sampling here is potentially due to the minimum train trip frequency requirement.
[3] Translated to international equivalents.

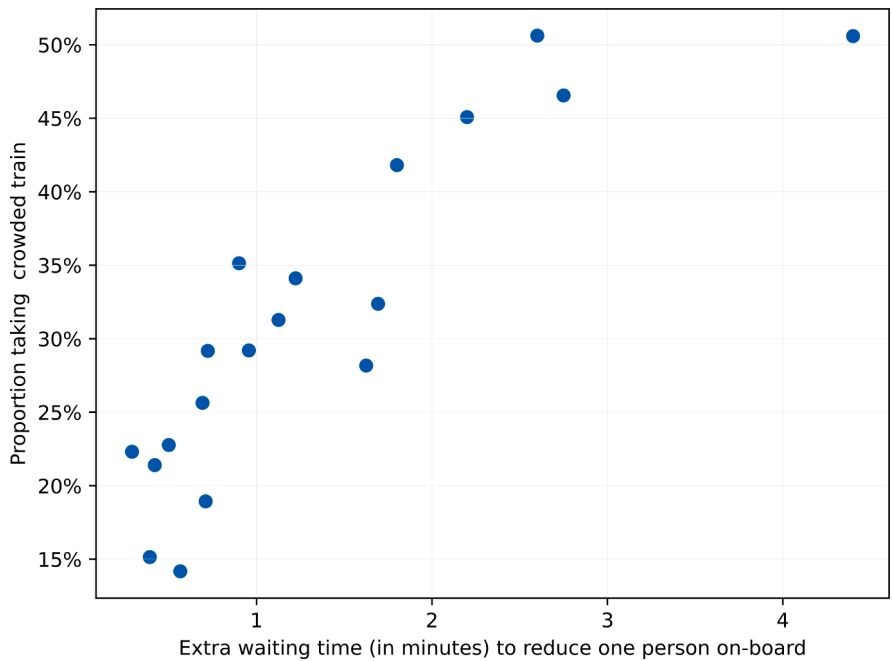

**Fig. 3.** Proportion of observations choosing crowded train versus extra waiting time to reduce one person on-board.

The class-specific taste parameters ($\beta_s$) and membership coefficients ($\gamma_{ks}$, $\delta_s$) are simultaneously estimated using PythonBiogeme (Bierlaire, 2016).

## 3. Results

Since the experiment was conducted in the context of the COVID-19 pandemic, one fear was that a large number of respondents would simply opt-out of using trains altogether. Ultimately, this was not the case and only about 4% of the respondents always opted-out while 13% never opted-out in the 15 situations they faced. A substantial number of respondents always ranked waiting higher than the crowded train (~13%) or vice versa (~3%). However, only about 1% and 3% always chose waiting or taking the crowded train as their top option, respectively. As expected the proportion of observations where the respondent takes the crowded train rises with the extra waiting time required to reduce one person on-board (Fig. 3).

We tried a number of utility specifications, in particular varying whether attributes were modelled as having a linear or non-linear effect. Since, perceived infection risk (and, therefore, disutility of a train alternative) may be higher when two contributing factors are higher together, we also included interaction effects of crowding with infection rate and in-vehicle times in the utility specification. Ultimately, the specification in Equation (6) within MNL was found to provide the most informative model parameters. Table 2 gives an overview of attributes included in the final model.

$$\begin{aligned}
V_{\text{train 1}} &= \beta^{\text{crowd: can sit alone}} \cdot crowd_{18,\text{ train 1}} + \beta^{\text{crowd: not crowded}} \cdot crowd_{23,\text{ train 1}} \\
&\quad + \beta^{\text{crowd: quite crowded}} \cdot crowd_{28,\text{ train 1}} + \beta^{\text{crowd: almost full}} \cdot crowd_{36,\text{ train 1}} \\
&\quad + \beta^{\text{WT}} \cdot WT_{\text{train 1}} + \beta^{\text{crowd} \times \text{infect}} \cdot crowd_{\text{train 1}} \cdot infect + \beta^{\text{crowd} \times \text{IVT}} \cdot crowd_{\text{train 1}} \cdot IVT \\
V_{\text{train 2}} &= \beta^{\text{crowd: can sit alone}} \cdot crowd_{18,\text{ train 2}} + \beta^{\text{crowd: not crowded}} \cdot crowd_{23,\text{ train 2}} \\
&\quad + \beta^{\text{crowd: quite crowded}} \cdot crowd_{28,\text{ train 2}} + \beta^{\text{crowd: almost full}} \cdot crowd_{36,\text{ train 2}} \\
&\quad + \beta^{\text{WT}} \cdot WT_{\text{train 2}} + \beta^{\text{crowd} \times \text{infect}} \cdot crowd_{\text{train 2}} \cdot infect + \beta^{\text{crowd} \times \text{IVT}} \cdot crowd_{\text{train 2}} \cdot IVT \\
V_{\text{opt-out}} &= \beta^{\text{opt-out}} + \beta^{\text{IVT}} \cdot IVT \\
&\quad + \beta^{\text{infect: 0.01}} \cdot infect_{0.01} + \beta^{\text{infect: 0.5}} \cdot infect_{0.5} + \beta^{\text{infect: 2}} \cdot infect_2 + \beta^{\text{infect: 10}} \cdot infect_{10}
\end{aligned} \quad (6)$$

The MNL model shown in Table 3 is finalized by removing insignificant ($p > 0.10$) parameters one-by-one. As shown, all parameters have the expected signs and magnitudes: the likelihood of choosing an option generally decreases with increasing crowding and waiting time while the likelihood of opting out increases with increasing infection rate. A small and large non-linear effect is found for the highest attribute levels of crowding and infection rate, respectively. Since respondents were only shown graphics for on-board crowding, meaningful non-linear effects possibly indicate that respondents may be applying subjective labels (Li and Hensher, 2013); for instance (as will be shown in the LCCM results) assigning a much higher utility to being able to sit alone than to a similar reduction in crowding otherwise. Although the coefficients for in-vehicle time and the interaction between crowding and in-vehicle time were not significant for the MNL model, we keep them in the specification tested for the LCCM.

We examined whether removing respondents who consistently ranked the crowded train higher or lower than the other train alternative (~16% of the sample) affected the results. The MNL model estimated from this reduced sample shows a slightly lower impact of crowding and infection rates, and a higher (negative) impact of opting out. A possible reason for this is that those who consistently ranked less crowded trains higher often selected opting out as their most preferred option. We also tested consistency in responses to check for respondent fatigue. To do this, we split the choice dataset into two—responses to the first eight choice situations and the last seven choice situations—and estimate the described MNL model for each set. While the value of crowding (relative to waiting time) remains similar, the model for the last seven situation shows larger and smaller impacts for infection rate and opting out, respectively. We suspect that as respondents gain experience with the choice situation, they are able to better discriminate between the contextual variable that is infection rate.

For the LCCM, we first find the optimal number of classes using an intercept-only class membership function. Typically, this is done using the model fit indicators, particularly the Bayesian information criterion (BIC), which explicitly penalize the number of parameters in the model. In our case, model fit indicators continued to improve as we increased the number of classes (we checked up to 6 classes). Therefore, we chose the 2-class model as it, in our opinion, best described heterogeneity in behaviour. While the 2-class model clearly delineated two behavioural types, adding more classes yielded intermediate groups without adding more insights. Moreover, adding more groups resulted in higher standard errors of estimated parameters and even led to unexpected parameter signs for higher number of classes. Next, the choice models in each class are finalized in the same way as the MNL model: by removing insignificant (p > 0.10) parameters one-by-one. Finally, all non-correlated observable individual characteristics (from section 2.1.2) are included in the class membership function and eliminated one-by-one if they are insignificant to arrive at the final model shown in Table 3.

**Table 2**
Overview of attributes included in the final choice model.

| Attributes | Symbol | Explanation | Range |
|---|---|---|---|
| **Choice coefficients** | | | |
| Crowding (level $i$) | $\beta^{\text{crowd: }i}$ | Categorical (effect coded) | |
| Waiting time | $\beta^{\text{WT}}$ | All time attributes are in minutes | 3–25 |
| In-vehicle time | $\beta^{\text{IVT}}$ | | 10–40 |
| Infection rate (level $i$) | $\beta^{\text{infect: }i}$ | Categorical (effect coded); percentage of population | 0.01–10 |
| Opt-out constant | $\beta^{\text{opt-out}}$ | | |
| **Personal characteristics** | | | |
| Age | $\beta^{\text{age}}$ | Ordinal in ascending order: | 1–7 |
| | | 18–24, 25–34, 35–44, 45–54, 55–64, 65–74, 75–84 | |
| Gender | $\beta^{\text{female}}$ | Categorical (effect coded): | |
| | | female, male | |
| Train use frequency during COVID | $\beta^{\text{train freq. covid}}$ | Ordinal in ascending order: | 1–4 |
| | | never, once per month, 1–3 times per month, >4 times per week | |

**Table 3**
Estimation results of the 2-class LCCM.

| Model | MNL | | | LCCM 2-Class | | | | | |
|---|---|---|---|---|---|---|---|---|---|
| # Parameters | 10 | | | 23 | | | | | |
| Initial LL | −8453.822 | | | −8054.030 | | | | | |
| Final LL | −7572.643 | | | −6498.217 | | | | | |
| Adjusted $\rho^2$ | 0.103 | | | 0.190 | | | | | |
| BIC | 15234.77 | | | 13202.246 | | | | | |
| | | | | *Class-specific choice models* | | | | | |
| | | | | **Class 1: COVID Conscious Travellers** | | | **Class 2: Infection Indifferent Travellers** | | |
| Class Size | | | | 53.73% | | | 46.27% | | |
| | Coeff. | p-val | Scaled | Coeff. | p-val | Scaled | Coeff. | p-val | Scaled |
| $\beta^{\text{crowd: almost empty}}$ | 1.317 | – | −42.76 | 2.230 | – | −159.29 | 0.690 | – | −18.16 |
| $\beta^{\text{crowd: can sit alone}}$ | 0.245 | 0.00 | −7.95 | 0.792 | 0.00 | −56.74 | 0 | – | – |
| $\beta^{\text{crowd: not crowded}}$ | −0.188 | 0.00 | 6.10 | −0.531 | 0.00 | 37.93 | 0.110 | 0.01 | −2.89 |
| $\beta^{\text{crowd: quite crowded}}$ | −0.558 | 0.00 | 18.12 | −0.921 | 0.00 | 65.79 | −0.262 | 0.00 | 6.89 |
| $\beta^{\text{crowd: almost full}}$ | −0.816 | 0.00 | 26.49 | −1.570 | 0.00 | 112.14 | −0.538 | 0.00 | 14.16 |
| $\beta^{\text{WT}}$ | −0.031 | 0.00 | 1 | −0.014 | 0.02 | 1 | −0.038 | 0.00 | 1 |
| $\beta^{\text{crowd}\times\text{infect}}$ | −0.0012 | 0.00 | 0.004 | −0.0046 | 0.00 | 0.33 | −0.0026 | 0.00 | 0.07 |
| $\beta^{\text{crowd}\times\text{IVT}}$ | – | – | – | – | – | – | – | – | – |
| $\beta^{\text{infect: 0.01}}$ | −0.397 | – | 12.89 | −0.720 | 0.00 | 51.43 | 0 | – | – |
| $\beta^{\text{infect: 0.1}}$ | 0.059 | – | −1.92 | −0.213 | – | 15.21 | −0.488 | – | 12.84 |
| $\beta^{\text{infect: 0.5}}$ | 0 | – | – | 0.133 | 0.08 | −9.5 | 0 | – | – |
| $\beta^{\text{infect: 2}}$ | 0.387 | 0.00 | −12.56 | 0.521 | 0.00 | −37.21 | 0.254 | 0.00 | −6.68 |
| $\beta^{\text{infect: 10}}$ | 0.375 | 0.00 | −12.18 | 0.279 | 0.05 | −19.93 | 0.234 | 0.00 | −6.16 |
| $\beta^{\text{IVT}}$ | – | – | – | – | – | – | – | – | – |
| $\beta^{\text{opt-out}}$ | −0.424 | 0.00 | 13.77 | 0.927 | 0.00 | −66.21 | −2.02 | 0.00 | 53.16 |
| | | | | *Class membership model* | | | | | |
| | | | | **Class 1: COVID Conscious Travellers** | | | **Class 2: Infection Indifferent Travellers** | | |
| | | | | Coeff. | p-val | Scaled | Coeff. | p-val | Scaled |
| $\beta^{\text{intercept}}$ | | | | 0 | – | – | −1.160 | 0.00 | 1 |
| $\beta^{\text{age}}$ | | | | – | – | – | −0.107 | 0.08 | 0.09 |
| $\beta^{\text{female}}$ | | | | – | – | – | −0.275 | 0.01 | 0.24 |
| $\beta^{\text{train freq. covid}}$ | | | | – | – | – | 0.820 | 0.00 | −0.71 |

The choice parameters in both classes have signs and magnitudes in line with expectations. Results show that, in general, higher levels of crowdedness, waiting times, and infection rates all reduce travellers' willingness to board a particular train alternative and increase the probability of opting out. Surprisingly, for the LCCM too, in-vehicle time—time to be spent in an enclosed train coach—does not affect travellers' decisions indicating that they might be underestimating the importance of duration of exposure on the risk of infection.

To enable comparisons of coefficients across models, we calculate the ratio of each attribute coefficient to the waiting time coefficient under the scaled values columns in Table 3. As can be seen from the scaled columns, the two estimated classes differ strongly on the relative impact of level of crowdedness and infection rates. Moreover, the general propensity to opt-out has a very large effect in both classes but is in opposite directions. Based on these differences, we call the first class 'COVID Conscious' as decisions in this group are more strongly driven by the level of crowdedness, infection rates, and the expected number of infected persons on-board (approximated by the interaction effect of crowdedness and infection rate). In contrast, the second class, which we call 'Infection Indifferent', is affected by these factors to a lesser degree and is also less likely to opt-out on average.

Since respondents are assigned to classes probabilistically, econometric indicators calculated for the two classes do not directly apply to individual respondents. Instead, they form the lower and upper boundaries for individual indicators which are obtained as the weighted average of the estimates for the two classes. Averaging the individual-specific indicators gives an aggregate value over the whole sample. The weights—posterior membership probabilities—are calculated by multiplying the prior probabilities (obtained by applying the class membership model) with the likelihood of observing individual respondents' sequence of choices and then normalizing (Hensher et al., 2015). The sum of the posterior probabilities for each class gives the class sizes shown in percentage in Table 3. To examine the extent to which individuals belong to either class, in Fig. 4, we plot the distribution of absolute difference between posterior class membership probabilities (higher values indicate a more deterministic assignment to either class). As can be seen, about 85% of respondents are assigned to one class with a probability of 95% or more. In the following, we discuss traits of the two latent classes—alternatively referring to prototypical (i.e., representative) travellers of either class—rather than focussing on aggregate estimates. Note that, given the posterior probability distribution, for most respondents, individual-specific estimates would be close to those calculated for one of the classes.

Typically, the effect of crowding has been modelled as an in-vehicle time multiplier (Li and Hensher, 2011; Wardman and Whelan, 2011). The idea being that the disutility of crowdedness should be larger for longer trips because passengers have to be in a crowded vehicle for a longer time. We included crowding in our model both as a constant penalty as well as an interaction effect with in-vehicle time (i.e., as a multiplier). As shown in Table 3, the time multiplier parameters were not significant. When constant penalties are excluded, the time multiplier parameters are significant but, similar to Kroes et al. (2014), the model has a significantly lower goodness of fit. We note that constant penalties may only be performing better for the range of in-vehicle times that were tested in this survey or because in-vehicle times were only included as context effects in the experiment.

Equation (7) shows how the values of crowding are calculated. Since we estimate parameters for each crowding level ($i$), the coefficient for change in crowding between two levels ($\beta^{\text{crowd}:i\to i+1}$) is given by the difference in the utility contributions divided by the difference in the number of persons on-board ($x^i$). This crowding coefficient divided by the coefficient for waiting time gives the value of crowding in terms of waiting time between those levels ($\Upsilon^{i\to i+1}$). The average value of crowding ($\Upsilon$) is given by the weighted average of these individual values of crowding.

$$\beta^{\text{crowd}:i\to i+1} = \frac{\beta^{\text{crowd}:i} - \beta^{\text{crowd}:i+1}}{x^i - x^{i+1}}$$

$$\Upsilon^{i\to i+1} = \frac{\beta^{\text{crowd}:i\to i+1}}{\beta^{\text{WT}}} \qquad (7)$$

$$\Upsilon = \frac{\Upsilon^{1\to 2}\cdot(x^2-x^1)+\Upsilon^{2\to 3}\cdot(x^3-x^2)+\Upsilon^{3\to 4}\cdot(x^4-x^3)+\Upsilon^{4\to 5}\cdot(x^5-x^4)}{(x^2-x^1)+(x^3-x^2)+(x^4-x^3)+(x^5-x^4)}$$

Prototypical travellers of the COVID Conscious class, are willing to wait an extra 8.75 min, on average, to reduce just one person on-board. As shown in Fig. 5, travellers in this class are willing to wait the most when there is a possibility to sit alone. This is indicative of the aversion towards infection risk in this class as well as the general framing of the choice situations in the context of the pandemic. Note that this is in contrast to previous studies which typically report that the impact of crowding increases with the number of persons on-board, especially after 60–80% load factor (Hörcher et al., 2017; Wardman and Whelan, 2011; Yap et al., 2018). The value of crowding for Infection Indifferent travellers seems to be more in line with values from previous studies (albeit on the higher end) with an average willingness to wait of 1.04 min to reduce one person on-board. (The average willingness to wait in the MNL model is 2.23 min per person.)

Previous studies that evaluate the effect of seating occupancies—either as constant penalties or in-vehicle time multipliers—and waiting times can be compared with our results. To convert coefficients from such studies to units comparable to ours, where required, we assume a total seat capacity of 40 and an in-vehicle time range of 10–40 min. Preston et al. (2017) observed stated choices in an experiment where respondents chose between two trains: the first which was due but with no possibility to sit and the second with a given expected waiting time and one of two lower crowding levels. They report constant crowding penalties that lead to values of crowding in the range of 0.52–0.96 min per person. From an experiment similar to ours, Kroes et al. (2013) find a willingness to wait of 0.15 and 0.33 min per person to reduce crowdedness from 75% to 50% and from 100% to 75%, respectively. Assuming expected waiting times to be half of headways, Tirachini et al. (2013) find similar values of 0.15–0.6 min per person from a mode choice experiment. For comparison, note that we found the COVID Conscious and Infection Indifferent classes willing to wait 8.75 and 1.04 min per person, respectively. Douglas and Karpouzis (2006) and Sahu et al. (2018) use similar stated choice experiments for Sydney and Mumbai. They find that travellers are willing to wait 1.88–7.52 min and 3.58–14.32 min, respectively, for an uncrowded seat over a crowded seat alternative. Assuming 'not crowded' and 'almost full' to be the corresponding categories in our model, we find an extremely high value of 74 min and a more moderate 17 min in the COVID Conscious and Infection Indifferent classes, respectively.

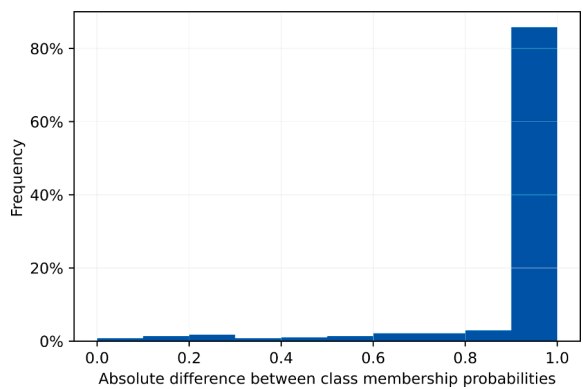

**Fig. 4.** Distribution of absolute difference between posterior class membership probabilities.

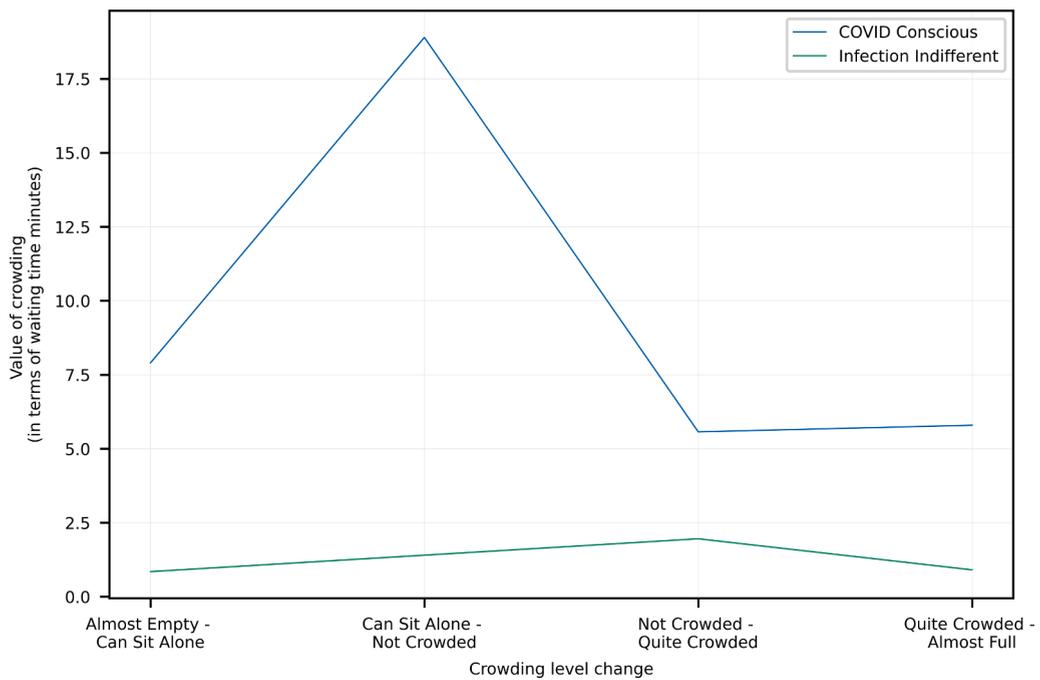

**Fig. 5.** Value of crowding (in terms of waiting time minutes per person on-board) between levels in the two traveller classes. The value of crowding is interpolated between can sit alone and not crowded in the Infection Indifferent class.

Using revealed preferences from smart card data in The Hague, however, Yap et al. (2018) found significantly lower values between 0.015 and 0.06 min per person (depending on in-vehicle times) for trams. Thus, while both the COVID Conscious and Infection Indifferent classes show higher values of crowding, the latter is much closer to pre-pandemic estimates from stated choice experiments.

As shown in Fig. 6, for both classes, the tendency to opt-out increases as a concave function of the prevalent infection rate. The effect plateaus at extreme infection rates (2% and 10%) indicating that travellers may be considering a threshold level beyond which the infection rate itself no longer contributes to perceived risk. A *t*-test could not reject the null hypothesis that the coefficients for these extreme infection rate levels are equal in either class (p-values: 0.17 and 0.91 for each class, respectively). Other consecutive levels are statistically different (p-value < 0.01). The graph for COVID Conscious prototypical travellers demonstrates, again, the strong preference to sit in an almost empty coach or to sit alone. Furthermore, note that the opt-out rates for crowded vehicles in the COVID Conscious class are fairly inelastic in relation to infection rates. This might indicate that travellers in this group would not feel safe travelling in crowded vehicles even when infection rates are back to pre-pandemic levels.

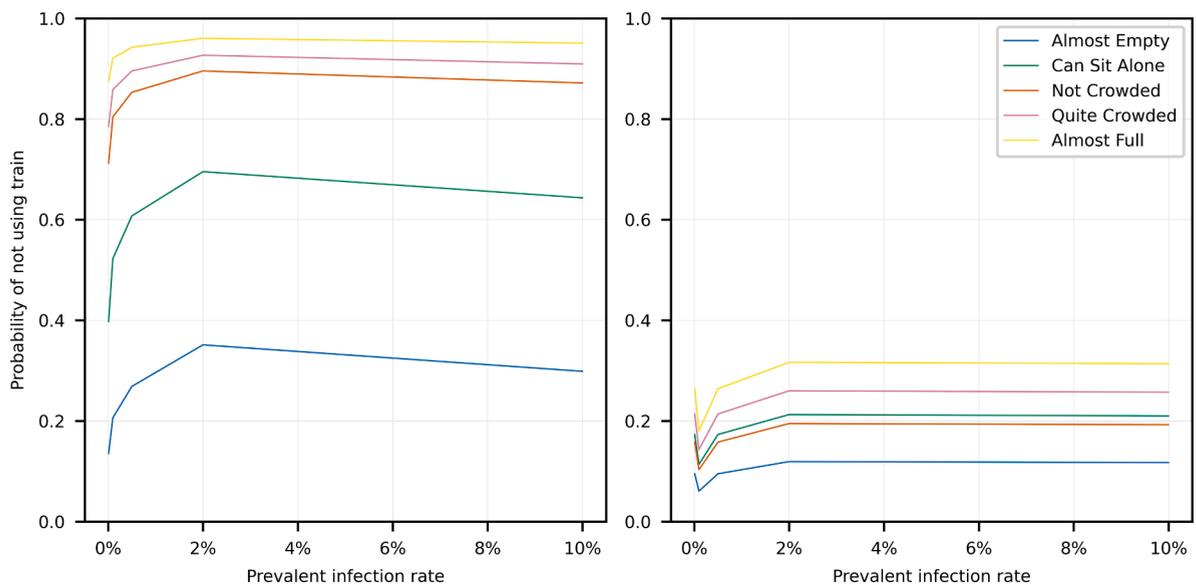

**Fig. 6.** Probability of not using trains versus infection levels for different crowding levels for the two classes (wait time for trains is 12 min). The dip at 0.1% for the Infection Indifferent class (right) is an artefact of the chosen reference level, arising because the two lowest infection rate levels have insignificant coefficients and are, consequently, fixed to zero.

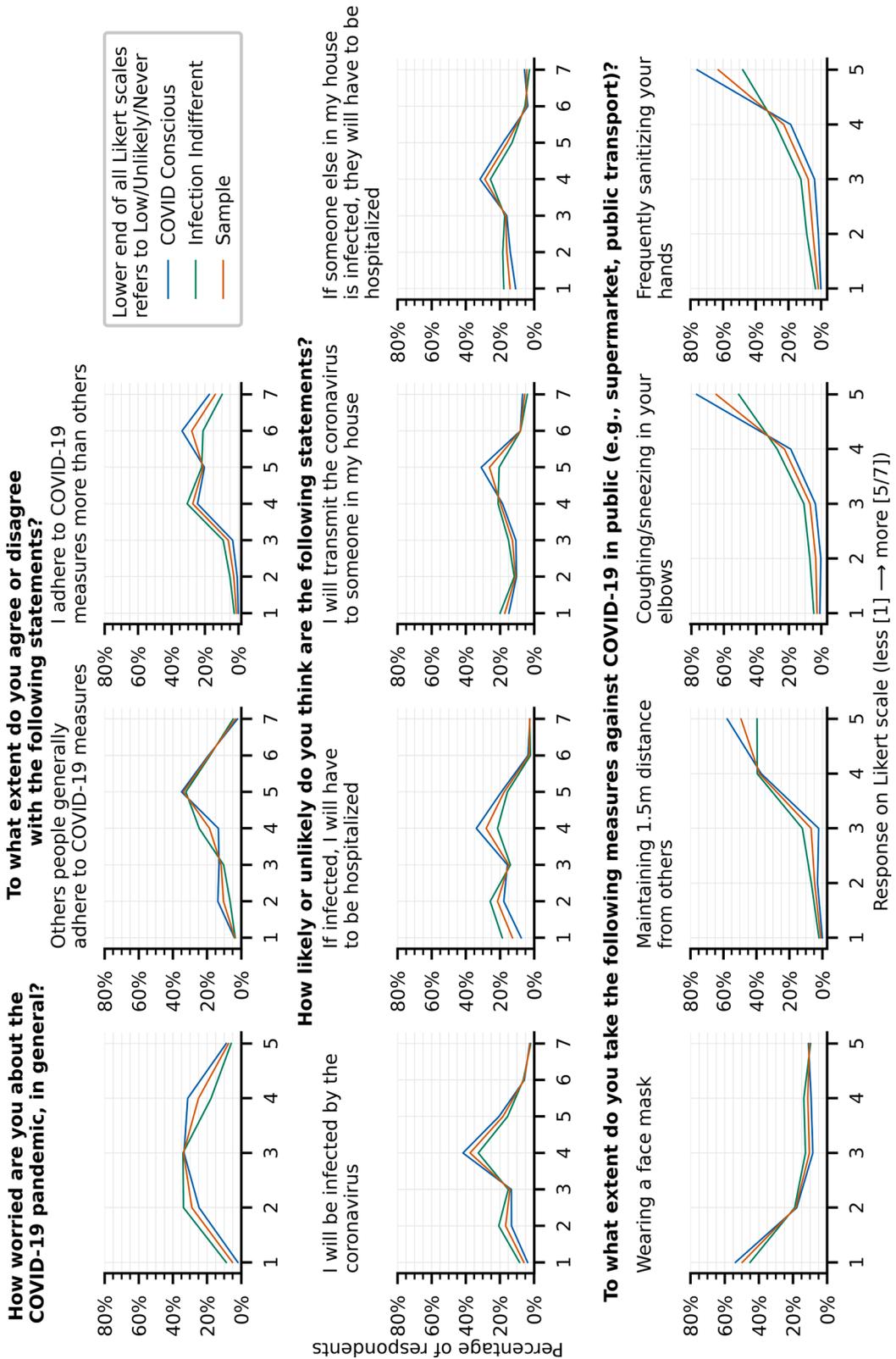

**Fig. 7.** Class profiles of personal, unobserved, qualitative measures related to the COVID-19 pandemic.

Amongst the individual characteristics collected, three variables contributed to explaining the differences in behaviour between the two classes. Older and female respondents were over-represented amongst COVID Conscious travellers whereas those reporting higher train use during the COVID-19 restrictions were likely to be Infection Indifferent. Presumably, older people, are more risk averse due to a higher vulnerability to the disease. While the disease does not seem to affect women more severely than men, female respondents have often been shown to be more risk averse in their decisions (e.g., de Palma and Picard (2005)). Kluwe-Schiavon et al. (2021) also find that older and female respondents had a lower COVID-19 risk tolerance for economic opportunity. The causal relationship between higher train use during the pandemic and lower risk aversion may be in either direction. Travellers with a higher probability of being Infection Indifferent may have used the train more frequently because they are not particularly averse to the COVID-19 risk. Conversely, lower risk aversion amongst those who use the train more frequently during the pandemic might be explained by the existence of the description-experience gap when evaluating risky choices. When judging the likelihood of contracting COVID-19 on public transport, these travellers may be depending more on their experience rather than the risk described by authorities (Barron and Erev, 2003). When people make decisions based on experience, they do not account for rare events as much as the objective probabilities of such events suggest they should (Hertwig and Erev, 2009). A little surprisingly, having the possibility to conduct the trip with a private mode (e.g., car, bicycle, walk) was not related to the propensity to opt-out of the train alternatives.

Fig. 7 shows the posterior distributions of some unobserved COVID-19-related qualitative measures. To obtain these distributions, the posterior membership probabilities are summed class-wise for all respondents who select a particular level for an indicator. The class-wise sums are then normalized by the respective class-sizes to find the percentage of each class that would select a given level on the indicator being analysed. Since respondents are only probabilistically assigned to either class, we note that the distributions convey the characteristics of the latent class rather than that of the individuals. Although the two classes do not differ too strongly on these factors, small differences can be noted. The COVID Conscious class tends to be more worried about the pandemic, specifically, about being hospitalized and spreading the infection to someone in their household. A moderate correlation exists between age and worrying about being hospitalized but not with other factors (older travellers are over-represented in this class). Additionally, travellers highly likely to be COVID Conscious reported themselves to be more rule-following, indicating that they followed advice such as frequently sanitizing hands and maintaining 1.5 m distances in public places. Moreover, they also had a more negative opinion about the degree to which others followed these rules. Indicative of the long drawn national debate over it (DutchNews.nl, 2020), face mask use is unpopular with both classes and largely uncorrelated with the degree to which other measures are followed.

## 4. Conclusion

The COVID-19 pandemic has had an extensive impact on public transport. As a result of actions and advisories aimed at containing the disease, public transport ridership has declined sharply, perceptions regarding this mode have become more negative, and there has been a shift to personal transport modes. Consequently, changes in traveller behaviour in order to minimize exposure to the virus are expected. Moreover, these changes may be sustained through different stages of the pandemic and even have a significant effect on public transport demand after the pandemic. While a number of studies examine current ridership patterns and anticipated transport preferences, few have investigated trade-offs in the age of COVID-19 via choice analysis in detail.

In this study, we analysed traveller behaviour related to factors affecting the risk of COVID-19 transmission in public transport with a stated choice experiment. Since one of the most important ways to avoid exposure is to reduce contact with other people, we measured travellers' (potentially updated) valuation for on-board crowding. To do this, we obtained respondents' preferences between a crowded-but-low-wait-time and a less-crowded-but-higher-wait-time (and an opt-out) alternative. Choices were presented in the context of exposure duration (operationalised as the in-vehicle time of the alternatives) and infection rate to examine the effects of these risk-contributing factors on choice behaviour.

Responses were collected from train travellers in the Netherlands at the end of the first infection wave (May 20–25, 2020), just as the first restrictions were being lifted and new regulations were setup for travel in public transport. We believe that behavioural insights from this study will contribute to better demand forecasting. In addition to providing insights regarding when travellers opt-out (i.e., choose not to travel), by modelling decisions of the type 'to board or to wait' we can provide an important behavioural input to agent-based models (e.g., Cats and Gkioulou (2017)) that commonly simulate this choice situation. These results will also be valuable in informing public transport policy decisions, not only in the current pandemic but also future ones.

Applying a latent class choice model, we found two, nearly equal-sized traveller segments: COVID Conscious and Infection Indifferent. While higher crowding levels and infection rates reduce the willingness to board a train in both, the effect of these factors is much larger in the COVID Conscious segment. Value of crowding, measured as the number of minutes travellers are willing to wait to reduce one person on-board, is also significantly higher in this class (on average 8.75 min/person) and increases sharply with the possibility to sit alone. In contrast, Infection Indifferent travellers' value of crowding (on average 1.04 min) is comparable to, although slightly on the higher end of, pre-pandemic evaluations. Moreover, unlike their counterparts, COVID Conscious travellers are highly affected by the prevalent infection rate, particularly at low crowding levels, and are more likely to opt-out in general. Surprisingly, neither group took the exposure duration into account. Older and female respondents are over-represented in the COVID Conscious class while those who report higher train use during the pandemic are more likely to belong to the Infection Indifferent class. Finally, distributions of COVID-19-related indicators showed that COVID Conscious travellers were more worried about the pandemic, considered themselves and household members more likely to be hospitalised if infected, and reported themselves to be following related measures to a higher degree.

Although the crowding valuation for the COVID Conscious class is high, we note that it does not necessarily imply that travellers are willing to accept hour-long waits. Firstly, respondents had the option to opt-out and secondly, applications of the choice model must, technically, be within the range of waiting times included in the experiment (3–25 min). Thus, the COVID Conscious class model, in fact, indicates a very strong preference for either selecting the less crowded train or opting out altogether. On the other hand, prototypical travellers of the Infection Indifferent class, show a more nuanced decisions between taking the crowded train or waiting. Furthermore, while we cannot rule out the possibility that the COVID Conscious class has an inherently lower value of time, the results as a whole strongly suggest that the high values of crowding are driven by sensitivity to pandemic risks. For instance, the COVID Conscious class has a higher likelihood of opting out and is impacted more strongly by infection rates. Moreover, it also scores higher on indicators such as worrying about COVID-19, likelihood of being hospitalized, hygiene-related rule following, etc.; all of which imply a strong impact of the pandemic.

A variety of direct and indirect effects of the pandemic have led public transport ridership to plummet. Given the importance of public transport in economic recovery and sustainable mobility, authorities and operators need to work to improve travellers' perception about public transport and slow down the shift to non-sustainable modes. Ridership levels have typically returned to normal at the end of previous localized catastrophic events, such as epidemics and security threats (Gkiotsalitis and Cats, 2020). However, the COVID-19 pandemic is unprecedented in its spatial and temporal scale with regions around the world going in and out of lockdowns over an extended period of time. Thus, authorities cannot depend on ridership to improve by itself but must actively work towards increasing public transport demand while providing this essential service safely.

The apprehension of the COVID Conscious segment seemingly follows calls from authorities to avoid public transport. While such calls are compatible with the intuition that sharing confined spaces may be unsafe, there is little to no hard evidence of outbreaks linked to public transport. This might suggest that public transport travel could be safe if recommended measures are implemented (Gkiotsalitis and Cats, 2020; Goldbaum, 2020; Schive, 2020; UITP, 2020). Yet, the fact that over two-thirds of COVID Conscious travellers in our sample are unwilling to travel if they cannot find an empty row, regardless of infection level, is an indication of how difficult it will be to restore travellers' confidence and foreshadows lingering behavioural adaptations from the pandemic in the future. Where trips cannot be replaced by telecommuting or active modes, providing crowding information for public transport can be the key. For low infection rates (0.1%), when there is a possibility to sit in an almost empty coach or with the adjacent seat empty, 50–80% of COVID Conscious travellers in our sample indicated that they will use the train. By highlighting which trains and coaches will be less crowded, these travellers can adjust their departure times, routes, and even the choice of which coach to board. Assuming that these travellers overestimate the likelihood of contracting COVID-19 on-board public transport, more experience with travelling (even in less crowded vehicles) could bring their assessments in line with reality. Future studies may also look into the effect of other risk mitigating actions, such as mask mandates and increased cleaning, as well as the (perceived) extent to which these are followed on travellers' risk perceptions.

While public transport may be safe with recommended measures, their overall lower concern regarding the virus and absence of substantial behavioural change, indicates that Infection Indifferent travellers may not be motivated to follow them carefully. Poor compliance from these travellers could increase the real as well as perceived risk for everyone and further drive the apprehension of other travellers. Therefore, we must continue to emphasize the need for simple measures such as face masks and recognize that returning to pre-pandemic levels of crowding while the prevalent infection levels are still significant would be reckless.

Although the stated choice experiment provides important behavioural estimates we note limitations arising from the information provided to respondents and the nature of such experiments. Firstly, the prevalent infection rates given as a contextual attribute can only be estimated, and trusted estimations may not be available everywhere. Even if they were available, one might question whether travellers actually consider this information directly or respond to more abstract cues, such as the intensity of regulations or media coverage. However, since it is difficult to recreate the entire context for the experiment, we used this single indicator (which is correlated to such cues). Furthermore, travellers were helped in anchoring the prevalent infection rates to abstract cues by providing the prevalent infection rates at two different dates. Nevertheless, care must be taken when transferring estimations to other situations. Future studies could focus on analysing the impacts of specific societal, political, and media cues on pandemic-related travel behaviour.

Secondly, crowding information is not commonly available although a growing number of public transport networks and trip planning applications now try to provide predictions in some format (see review in (Drabicki et al., 2020)). In their smartphone application, the Dutch railways show a (qualitative) three-level crowding indicator for most trains and a more precise 'seat-finder' on some trains, showing seat availability in different coaches (Nederlandse Spoorwegen, 2021). By prominently displaying crowding information, we may have drawn respondents' attention to this aspect more than usual, leading to higher crowding valuation. Previous studies (e.g., Yap et al. (2018)) have also claimed that travellers tend to demonstrate a higher value of crowding in stated choice experiments than is observed from passively collected data. Moreover, while we presented crowding (and waiting time) information as objectively true, in real life, depending on factors such as trust in the information provided by the operator, travellers may consider attributes of the second train to be uncertain and therefore attach a higher disutility to it. We believe reliable crowding information could be key in regaining travellers trust. Since, crowding is likely to be affected by subjective perceptions, studies assessing travellers' responses to different presentation formats and reliability levels would be critical for such developments.

Thirdly, we note that choices observed here are hypothetical and the situations do not directly reflect the various constraints arising from societal positions of individuals. These constraints have been previously found to play a significant role in travellers actual capacity to change behaviour (Kim et al., 2017). Thus, although we find an intent to avoid trains amongst COVID Conscious travellers their ability to do so may be limited because of employer constraints. In contrast, some Infection Indifferent travellers may indicate intent to use the trains but do not actually do so as they can work/study from home. We attempted to control for such factors by marking the opt-out option as 'I will not make this trip by train' and asking respondents if they had alternative modes for the stated trip purpose. We also asked respondents' family income range and (for working/studying individuals) how effectively they could work/study at home as well as the frequency of doing so. None of these variables were found to contribute towards explaining choice behaviour. Nevertheless, the precise distribution of travellers between the different classes and their propensity to opt-out must be used with care in applications, accounting for other individual constraints that might affect behaviour. We can, however, confidently interpret crowding valuations and the existence of significant risk-averse and indifferent traveller segments.

Finally, we stress that we have observed a snapshot of behaviour (and intent) for present and future circumstances. Since the global outbreak of the COVID-19 pandemic in March 2020, the situation has developed quickly and unpredictability. Given the widespread and extended impact of this pandemic, people have been rapidly adopting and changing behaviours for the evolving new realities encountered during the course of this crisis. Yet, these acquired behaviours also fluctuate as the level of precaution changes depending on a number of factors, such as local infection rates, vaccination status, personal impact assessment, and 'pandemic fatigue'. Thus, the trade-offs estimated here may change with new and significant developments; for instance, if a cure for COVID-19 emerges.

However, we emphasise that this does not obviate the utility of our findings and policy recommendations (or those of similar studies) for two reasons. First, given the uncertainty surrounding COVID-19, it has become almost impossible to confidently predict the 'end' of the pandemic and a return to a state of stability. With the emergence of new variants and questions surrounding the duration of vaccine protection and efficacy against new variants, we cannot rule out a return to a situation similar to the one analysed here. Second, and more importantly, our analysis contributes to a larger picture of traveller preferences (and subsequent policy recommendations) in key stages of the pandemic. In time, a *meta*-analyses charting traveller preferences over the pandemic may also be recommended. What we learn about travel behaviour from this pandemic will be instrumental in supporting policy-makers to act proactively in the next one.

**CRediT authorship contribution statement**

**Sanmay Shelat:** Conceptualization, Investigation, Formal analysis, Writing – original draft. **Oded Cats:** Conceptualization, Writing – review & editing, Supervision, Funding acquisition. **Sander van Cranenburgh:** Conceptualization, Writing – review & editing.

**Declaration of Competing Interest**

The authors declare that they have no known competing financial interests or personal relationships that could have appeared to influence the work reported in this paper.

**Acknowledgements**

Funding: This work was supported by the My-TRAC (H2020 Grant No. 777640) project and the Transport Institute of TU Delft.

**References**

Aaditya, B.h., Rahul, T.M., 2021. Psychological impacts of COVID-19 pandemic on the mode choice behaviour: A hybrid choice modelling approach. Transp. Policy 108, 47–58.
Aghabayk, K., Esmailpour, J., Shiwakoti, N., 2021. Effects of COVID-19 on rail passengers' crowding perceptions. Transport. Res. Part A Policy Pract. 154, 186–202.
Awad-Núñez, S., Julio, R., Gomez, J., Moya-Gómez, B., González, J.S., 2021. Post-COVID-19 travel behaviour patterns: impact on the willingness to pay of users of public transport and shared mobility services in Spain. Eur. Transport Res. Rev. 13, 20.
Barron, G., Erev, I., 2003. Small feedback-based decisions and their limited correspondence to description-based decisions. J. Behav. Decis. Making 16 (3), 215–233.
Beck, M.J., Hensher, D.A., 2020. Insights into the impact of COVID-19 on household travel and activities in Australia – The early days of easing restrictions. Transp. Policy 99, 95–119.
Bierlaire, M., 2016. PythonBiogeme: a short introduction, *Series on Biogeme*. Transport and Mobility Laboratory, School of Architecture, Civil and Environmental Engineering, Ecole Polytechnique Fédérale de Lausanne, Switzerland.
Bucsky, P., 2020. Modal share changes due to COVID-19: The case of Budapest. Transport. Res. Interdisciplinary Perspectives 8, 100141. https://doi.org/10.1016/j.trip.2020.100141.
Cats, O., Gkioulou, Z., 2017. Modeling the impacts of public transport reliability and travel information on passengers' waiting-time uncertainty. EURO J. Transport. Logist. 6 (3), 247–270.
Centraal Bureau voor de Statistiek, 2020. StatLine.
Cho, S.-H., Park, H.-C., 2021. Exploring the Behaviour Change of Crowding Impedance on Public Transit due to COVID-19 Pandemic: Before and After Comparison. Transport. Lett. 13 (5-6), 367–374.
Citymapper, Citymapper Mobility Index. Citymapper.com/CMI. Accessed on: 20 February 2021.
de Palma, A., Picard, N., 2005. Route choice decision under travel time uncertainty. Transport. Res. Part A Policy Pract. 39 (4), 295–324.
Department for Transport, U.K., Coronavirus (COVID-19): safer travel guidance for passengers. https://www.gov.uk/guidance/coronavirus-covid-19-safer-travelguidance-for-passengers. Accessed on: 4 December 2020.
Douglas, N., Karpouzis, G., 2006. Estimating the passenger cost of train overcrowding, *29th Australian Transport Research Forum*, Gold Coast, Queensland, Australia.
Drabicki, A., Kucharski, R., Cats, O., Szarata, A., 2020. Modelling the effects of real-time crowding information in urban public transport systems. Transportmetrica A: Transport Science 17 (4), 675–713.
Despite the Government U-Turn, the Dutch Are Still Unwilling to Wear Masks. Updated October 2, 2020, 2020, accessed 20 February 2021, https://www.dutchnews.nl/news/2020/10/despite-the-government-u-turn-the-dutch-are-still-unwilling-to-wear-masks/.
Gerhold, L., 2020. COVID-19: Risk Perception and Coping Strategies. 10.31234/osf.io/xmpk4.

Gkiotsalitis, K., Cats, O., 2020. Public transport planning adaption under the COVID-19 pandemic crisis: literature review of research needs and directions. Transport Rev. 41 (3), 374–392.
Goldbaum, C., 2020. Is the Subway Risky? It May Be Safer Than You Think, *The New York Times*. The New York Times.
Google LLC, Google COVID-19 Community Mobility Reports. https://www.google.com/covid19/mobility/. Accessed on: 20 February 2021.
Hensher, D.A., Balbontin, C., Beck, M.J., Wei, E., 2022. The impact of working from home on modal commuting choice response during COVID-19: Implications for two metropolitan areas in Australia. Transport. Res. Part A Policy Pract. 155, 179–201.
Hensher, D.A., Rose, J.M., Collins, A.T., 2011. Identifying commuter preferences for existing modes and a proposed Metro in Sydney, Australia with special reference to crowding. Public Transport 3 (2), 109–147.
Hensher, D.A., Rose, J.M., Greene, W.H., 2015. Applied Choice Analysis, 2nd ed. Cambridge University Press, Cambridge.
Hertwig, R., Erev, I., 2009. The description–experience gap in risky choice. Trends in Cognitive Sciences 13 (12), 517–523.
Hörcher, D., Graham, D.J., Anderson, R.J., 2017. Crowding cost estimation with large scale smart card and vehicle location data. Transport. Res. Part B: Methodol. 95, 105–125.
Hu, M., Lin, H., Wang, J., Xu, C., Tatem, A.J., Meng, B., Zhang, X., Liu, Y., Wang, P., Wu, G., Xie, H., Lai, S., 2020. Risk of Coronavirus Disease 2019 Transmission in Train Passengers: an Epidemiological and Modeling Study. Clin. Infect. Dis. 72 (4), 604–610.
Kim, C., Cheon, S.H., Choi, K., Joh, C.-H., Lee, H.-J., 2017. Exposure to fear: Changes in travel behavior during MERS outbreak in Seoul. KSCE J. Civ. Eng. 21 (7), 2888–2895.
Kluwe-Schiavon, B., Viola, T.W., Bandinelli, L.P., Castro, S.C.C., Kristensen, C.H., Costa da Costa, J., Grassi-Oliveira, R., Brañas-Garza, P., 2021. A behavioral economic risk aversion experiment in the context of the COVID-19 pandemic. PLoS ONE 16 (1), e0245261. https://doi.org/10.1371/journal.pone.0245261.
Kroes, E., Kouwenhoven, M., Debrincat, L., Pauget, N., 2013. On the Value of Crowding in Public Transport for Île-de-France, *OECD ITF Meeting on Valuing Convenience in Public Transport*, Paris.
Kroes, E., Kouwenhoven, M., Debrincat, L., Pauget, N., 2014. Value of Crowding on Public Transport in île-de-France, France. 2417, 37–45.
Lau, J.T.F., Yang, X., Tsui, H., Kim, J.H., 2003. Monitoring community responses to the SARS epidemic in Hong Kong: from day 10 to day 62. 57, 864–870.
Li, Z., Hensher, D., 2013. Crowding in public transport: a review of objective and subjective measures. J. Public Transport. 16 (2), 107–134.
Li, Z., Hensher, D.A., 2011. Crowding and public transport: A review of willingness to pay evidence and its relevance in project appraisal. Transp. Policy 18 (6), 880–887.
Nederlandse Spoorwegen, Travelling in Crowded Trains (in Dutch). https://www.ns.nl/reisinformatie/service-verbeteren/wat-kunt-je-doen-bij-drukte.html. Accessed on: 17 March 2021.
Openbaar Vervoer Nederland, Over het coronavirus (in Dutch). https://www.ov-nl.nl/over-het-coronavirus/. Accessed on: 4 December 2020.
Openbaar Vervoer Nederland, Welkom terug; OV-bedrijven staan klaar om reizigers te ontvangen (in Dutch). https://www.ov-nl.nl/welkom-terug-ov-bedrijven-staan-klaar-om-reizigers-te-ontvangen/. Accessed on: 4 December 2020.
Prather, K.A., Wang, C.C., Schooley, R.T., 2020. Reducing transmission of SARS-CoV-2. Science 368 (6498), 1422–1424.
Preston, J., Pritchard, J., Waterson, B., 2017. Train Overcrowding:Investigation of the Provision of Better Information to Mitigate the Issues. 2649, 1–8.
Rijksoverheid, 2020, Corona-aanpak: de volgende stap. https://www.rijksoverheid.nl/actueel/nieuws/2020/05/19/corona-aanpak-de-volgende-stap. Accessed on: 20 February 2021.
Rijksoverheid, Number of Infectious Persons (in Dutch). https://coronadashboard.rijksoverheid.nl/landelijk/besmettelijke-mensen. Accessed on: 20 February 2021.
Rubin, G.J., Amlôt, R., Page, L., Wessely, S., 2009. Public perceptions, anxiety, and behaviour change in relation to the swine flu outbreak: cross sectional telephone survey. 339, b2651.
Sahu, P.K., Sharma, G., Guharoy, A., 2018. Commuter travel cost estimation at different levels of crowding in a suburban rail system: a case study of Mumbai. Public Transport 10 (3), 379–398.
Schive, K., How safe is public transportation? https://medical.mit.edu/covid-19-updates/2020/09/how-safe-public-transportation. Accessed on: 9 March 2020.
Scorrano, M., Danielis, R., 2021. Active mobility in an Italian city: Mode choice determinants and attitudes before and during the Covid-19 emergency. Res. Transport. Econom. 86, 101031. https://doi.org/10.1016/j.retrec.2021.101031.
Tirachini, A., Cats, O., 2020. COVID-19 and Public Transportation: Current Assessment, Prospects, and Research Needs. JPT 22 (1). https://doi.org/10.5038/2375-0901.
Tirachini, A., Hensher, D.A., Rose, J.M., 2013. Crowding in public transport systems: Effects on users, operation and implications for the estimation of demand. Transport. Res. Part A: Policy Pract. 53, 36–52.
UITP, 2020. Public Transport is COVID-Safe.
Union Home Secretary, 2020. Guidelines on the measures to be taken by Ministries/Departments of Government of India, State/Union Territory Governments and Authorities for containment of COVID-19 Epidemic in the Country (24.03.2020), in: Affairs, M.o.H. (Ed.). Ministry of Home Affairs - Government of India, India.
Walker, J.L., Wang, Y., Thorhauge, M., Ben-Akiva, M., 2018. D-efficient or deficient? A robustness analysis of stated choice experimental designs. Theor. Decis. 84 (2), 215–238.
Wardman, M., Whelan, G., 2011. Twenty Years of Rail Crowding Valuation Studies: Evidence and Lessons from British Experience. Transport Rev. 31 (3), 379–398.
Yap, M., Cats, O., van Arem, B., 2018. Crowding valuation in urban tram and bus transportation based on smart card data. Transportmetrica A: Transport Sci. 16 (1), 23–42.